\documentclass[floatfix, preprint, onecolumn, prd,
showpacs, showkeys, preprintnumbers, nofootinbib, superscriptaddress]{revtex4-1}
\usepackage{natbib}
\usepackage{times}
\usepackage{amssymb,amsbsy,amsmath,amsfonts}
\usepackage{graphicx}
\usepackage{float}
\usepackage{color}
\usepackage{morefloats}
\usepackage{rotating}
\usepackage{srcltx}
\usepackage{slashed}
\usepackage{subfigure}
\usepackage{multirow}
\usepackage{verbatim}
\usepackage{tabularx}

\begin{document}

\title{$S$-wave $KK^*$ interaction in a finite volume and the $f_1(1285)$}

\author{Li-Sheng Geng}
\email[E-mail me at: ]{lisheng.geng@buaa.edu.cn}
\affiliation{School of Physics and Nuclear Energy Engineering, Beihang University, Beijing 100191, China}
\affiliation{International Research Center for Nuclei and Particles in the Cosmos, Beihang University, Beijing 100191, China}

\author{Xiu-Lei Ren}
\affiliation{School of Physics and Nuclear Energy Engineering, Beihang University, Beijing 100191, China}
\affiliation{Institut de Physique Nucl\'{e}aire, IN2P3-CNRS and Universit\'{e} Paris-Sud,
F-91406 Orsay Cedex, France}

\author{Yu Zhou}
\affiliation{School of Physics and Nuclear Energy Engineering, Beihang University, Beijing 100191, China}

\author{Hua-Xing Chen}

\affiliation{School of Physics and Nuclear Energy Engineering, Beihang University, Beijing 100191, China}
\affiliation{International Research Center for Nuclei and Particles in the Cosmos, Beihang University, Beijing 100191, China}

\author{Eulogio Oset}
\affiliation{Departamento de F\'{\i}sica Te\'orica and IFIC,
Centro Mixto Universidad de Valencia-CSIC,
Institutos de Investigaci\'on de Paterna, Aptdo. 22085, 46071 Valencia, Spain}
\date{\today}

\begin{abstract}
  Lattice QCD simulations provide a promising way to disentangle different interpretations of 
  hadronic resonances, which might be of particular relevance to understand the
  nature of the so-called $XYZ$ particles. Recent studies have shown that in addition to the well-established naive quark model picture,
  the axial-vector meson $f_1(1285)$ can also be understood as a dynamically generated state built upon
  the $KK^*$ interaction. In this work, we calculate the energy levels of the $KK^*$ system in the $f_1(1285)$ channel in finite volume using the chiral unitary approach.
  We propose to calculate the loop function in the dimensional regularization scheme, which is
  equivalent to the hybrid approach adopted in previous studies.  We also study the inverse problem of extracting the bound state information from synthetic lattice QCD data and
  comment on the difference between our approach and the L{\" u}scher method.
\end{abstract}

\maketitle

\section{Introduction}
The $f_1(1285)$ is a $I^G(J^{PC})=0^+(1^{++})$ axial-vector state with mass $m=1281.9\pm0.5$ MeV and width $\Gamma=24.2\pm1.1$  MeV \cite{Agashe:2014kda}. 
In the naive quark model, this state is assigned as a ${}^{2S+1}L_J={}^3 P_1$ state. In recent years, however, it has been suggested to be
a dynamically generated state  made from the $KK^*$ interaction, together with its axial-vector counterparts~\cite{ Lutz:2003fm,Roca:2005nm}. Such a picture has been extensively tested  in the past decade~\cite{Geng:2006yb, Roca:2006am,Lutz:2008km,Nagahiro:2008zza,Nagahiro:2008cv,Geng:2008ag,Nagahiro:2011jn,Roca:2012rx,  Zhou:2014ila,Aceti:2015zva,Aceti:2015pma}. All these studies yield consistent results that the ground-state axial-vector mesons can be understood as dynamically generated states or at least contain large pseudoscalar meson-vector meson components.

Lattice QCD (LQCD) simulations can be applied to study the properties of hadrons from first principles using quark and gluon degrees of freedom.  Although studies of ground-state hadrons, which do not decay via strong interactions, have been well established and have turned out to be quite successful~(see, e.g., Refs.~\cite{Fodor:2012gf,Aoki:2013ldr}), studies of resonances are more challenging, since they do not correspond to discrete energy levels on the lattice, and considerable additional efforts are needed to extract physical information from LQCD simulations. 
The L{\"u}scher method is the de facto standard one in the case of single channel two-body elastic scattering~\cite{Luscher:1986pf,Luscher:1990ux}. In this framework,
the discrete energy levels obtained in LQCD simulations are related to the scattering phase shifts in infinite spacetime.\footnote{Although in the present work we only need
to tackle a single channel problem, it should  be noted that the L{\"u}scher method has been generalized to the case of multichannel scattering~\cite{Liu:2005kr,Lage:2009zv,Bernard:2010fp,Hansen:2012tf}. A thorough study of the coupled $\pi K$ and $ \eta K$ channels has recently  been 
done in Refs.~\cite{Dudek:2014qha,Wilson:2014cna}  and for the coupled $\pi \pi$ and $K \bar{K}$ channels 
in Ref.~\cite{Wilson:2015dqa}. }
In Ref.~\cite{Doring:2011vk}, the authors have developed a new effective approach to connect the LQCD discrete energy levels with the physical phase shifts (energies) by keeping the full relativistic two-body propagator, from which the L{\"u}scher formulation can be derived. This new approach has been applied to study finite volume effects in the meson-baryon interaction in the J{\"u}lich model~\cite{Doring:2011ip}; the $K D$, $\eta D_s$ interaction~\cite{MartinezTorres:2011pr,Torres:2014vna};
the pion-kaon scattering~\cite{Doring:2011nd,Zhou:2014ana};
the $DN$, $\pi \Sigma_c$ interaction~\cite{Xie:2012np}; the $\pi\rho$ interaction~\cite{Roca:2012rx}; the $\pi\pi$ interaction~\cite{Chen:2012rp};  and  the $\bar{K}N$ interaction~\cite{MartinezTorres:2012yi}.

In the present work, we apply this approach to study the $KK^*$ interaction in the $f_1(1285)$ channel.  The $f_1(1285)$ is peculiar in the chiral unitary approach since it is made from the single channel $KK^*$ interaction and is located below the $KK^*$ threshold. As a result, it appears as a bound state in 
the dynamical picture.  Its experimental width can be obtained from considering other coupled channels (see, e.g., Ref.~\cite{Aceti:2015zva}) without affecting its nature being dominantly a $KK^*$ bound state.  Inclusion of high-order kernels in the chiral unitary approach  is found to have negligible effects on this picture~\cite{Zhou:2014ila}.

\section{Theoretical framework}

\subsection{Continuum}
 In the chiral unitary approach of Ref.~\cite{Roca:2005nm}, the $f_1(1285)$ is made of a single channel $\frac{1}{\sqrt{2}}(|\bar{K}^*K \rangle+ |K^*\bar{K})\rangle$. The  relevant $V$-matrix is
\begin{equation}\label{V matrix}
  V(s)=-\frac{\epsilon\cdot\epsilon'}{8f^2}(-3)\left[3s-(M^2+m^2+M'^2+m'^2)-\frac{1}{s}(M^2-m^2)(M'^2-m'^2)\right],
\end{equation}
where $f$ is the pseudoscalar decay constant, $s$ the invariant mass squared, $\epsilon$ ($\epsilon'$) stands for the polarization four-vector of the incoming (outgoing) $K^*$. The masses $M$ ($M'$), $m$ ($m'$) correspond to the initial (final) $K^*$ and $K$, respectively.  The potential $V$ is unitarized via the following Bethe-Salpeter equation~\cite{Roca:2005nm}:
\begin{equation}\label{BS equation}
  T=[1+V\hat{G}]^{-1}(-V)\vec{\epsilon}\cdot\vec{\epsilon}~',
\end{equation}
where $\hat{G}=G(1+\frac{1}{3}\frac{q^2}{M^2})$ and $q$ is given by
\begin{equation}\label{q_l}
  q=\frac{1}{2\sqrt{s}}\sqrt{[s-(M+m)^2][s-(M-m)^2]}.
\end{equation}
The  scalar loop function $G$ has the following form:
\begin{equation}\label{loop function}
  G(\sqrt{s})=i\int\frac{d^4q}{(2\pi)^4}\frac{1}{(P-q)^2-M^2+i\epsilon}\frac{1}{q^2-m^2+i\epsilon},
\end{equation}
with $P$ the total incident momentum, which in the center-of-mass frame is $(\sqrt{s},0,0,0)$.

The loop function $G$ is divergent and needs to be regularized. This can be done
either in the dimensional regularization scheme or in the cutoff scheme. In the former, the loop function reads
\begin{equation}\label{dimensional}
  \begin{split}
    G^D(\sqrt{s})=&\frac{1}{16\pi^2}\{a(\mu)+\textrm{ln}\frac{M^2}{\mu^2}+\frac{m^2-M^2+s}{2s}\textrm{ln}\frac{m^2}{M^2}\\ &+\frac{q}{\sqrt{s}}[\textrm{ln}(s-(M^2-m^2)+2q\sqrt{s})+\textrm{ln}(s+(M^2-m^2)+2q\sqrt{s})\\ &-\textrm{ln}(-s+(M^2-m^2)+2q\sqrt{s})-\textrm{ln}(-s-(M^2-m^2)+2q\sqrt{s})]\}.
  \end{split}
\end{equation}
In our work, the regularization parameters are chosen to be $ a(\mu)=-1.85$ and $ \mu =900$ MeV~\cite{Roca:2005nm}.

\subsection{Finite volume}
To study the $f_1(1285)$ meson in finite volume, one replaces $T$ of Eq.~(\ref{BS equation}) by $\tilde{T}$, obtained using the same equation with the same potential and replacing  the $G$-function in Eq.(\ref{BS equation}) by its counterpart  defined in a finite box of size $L$. The function
$G$ in finite volume, $\tilde{G}$, can be calculated again either in the dimensional regularization scheme, the cutoff scheme~\cite{Doring:2011vk}, or a combination of both---the hybrid approach~\cite{MartinezTorres:2011pr}. To remove small unphysical discontinuities  in the cutoff scheme, a smooth cutoff has been implemented in Ref.~\cite{Doring:2011vk}. In the hybrid approach~\cite{MartinezTorres:2011pr}, an average of the results obtained with several sharp cutoffs is taken.  This can save  computational time when very large cutoff values are used.

In principle in finite volume one mixes partial waves due to the cubic, rather than spherical, symmetry of the finite boxes chosen in the lattice simulations.
The problem has been thoroughly studied in Ref.~\cite{Luscher:1990ux} and it is particularly relevant when one performs lattice simulations for particles in a moving frame~\cite{Doring:2012eu,Rummukainen:1995vs,Kim:2005gf,Bour:2011ef,Beane:2011sc,Davoudi:2011md,Fu:2011xz,Leskovec:2012gb,Dudek:2012gj,Hansen:2012tf,Briceno:2012yi}.
The formulation for moving frames along the lines of Ref.~\cite{Doring:2011vk} is also done in Ref.~\cite{Doring:2012eu}. In the present paper we only study systems with the two particles at rest interacting
with $S$-waves. 
We shall discuss the mixing in detail in Sec. IV, but we anticipate that for the levels that we consider in the inverse analysis, only the single channel with L=0 is relevant.

In this work, we propose to calculate $\tilde{G}$ in the dimensional regularization scheme. Introducing the so-called
finite-volume correction, $\delta G$, $\tilde{G}$ can be written as:
\begin{equation}
\tilde{G}=G^D+\delta G,
\end{equation}
For the loop function of Eq.~(\ref{loop function}) , $\delta G$  has the following form~\cite{Geng:2011wq}:
\begin{equation}\label{delta-G-B}
  \delta G\equiv G(L)-G(\infty)=-\frac{1}{4}\int^1_0 dx \delta_{3/2}(\mathcal{M}^2(s)),
\end{equation}
where
\begin{equation}
  \begin{split}
    \mathcal{M}^2(s)=(x^2-x)s+ x M^2+(1-x) m^2-i\epsilon.
  \end{split}
\end{equation}
Depending on the value of $\sqrt{s}=\sqrt{P^2}$, $\tilde{G}$ needs to be treated differently. In the case of $\sqrt{s}>M+m$, $\delta_r(\mathcal{M}^2(s))$ can be written as a sum of the following three parts \cite{Bernard:2007cm,Ren:2013oaa}:
\begin{equation}
  \delta_r(\mathcal{M}^2(s))=g^r_1-g^r_2+g^r_3,
\end{equation}
where the $g^r_{1,2,3}$ are defined as
\begin{eqnarray}\label{gr1}\label{eq:g1r}
  g^r_1&=&\frac{1}{L^3}\sum_{\vec{q}}\left\{\frac{1}{[\frac{4\pi^2\vec{n}^2}{L^2}+\mathcal{M}^2(s)]^r}
  -\frac{1}{[\frac{4\pi^2\vec{n}^2}{L^2}+\mathcal{M}^2(m_{ss}^2)]^r}
  +\frac{r(x^2-x)(s-m_{ss}^2)}{[\frac{4\pi^2\vec{n}^2}{L^2}+\mathcal{M}^2(m_{ss}^2)]^{r+1}}\right\},\\
  g^r_2&=&\int_0^{+\infty}\frac{q^2 dq}{2\pi^2}\left\{\frac{1}{[\vec{q}^2+\mathcal{M}^2(s)]^r}
  -\frac{1}{[\vec{q}^2+\mathcal{M}^2(m_{ss}^2)]^r}
  +\frac{r(x^2-x)(s-m_{ss}^2)}{[\vec{q}^2+\mathcal{M}^2(m_{ss}^2)]^{r+1}}\right\},\\
  g^r_3&=&\delta_r(\mathcal{M}^2(m_{ss}^2))-r(x^2-x)(s-m_{ss}^2)\delta_{r+1}(\mathcal{M}^2(m_{ss}^2)).
\end{eqnarray}
The separation scale $m_{ss}$ needs to satisfy $m_{ss}<M+m=M_{K^*}+m_K$. In the case of $\sqrt{s}<M+m$, $\delta_r(\mathcal{M}^2)$ has a much simpler form~\cite{Geng:2011wq}:
\begin{equation}\label{eq:deltar}
  \delta_r(\mathcal{M}^2(s))=\frac{2^{-1/2-r}(\sqrt{\mathcal{M}^2})^{3-2r}}{\pi^{3/2}\Gamma(r)}\sum_{\vec{n}\neq 0}(L\sqrt{\mathcal{M}^2}|\vec{n}|)^{-3/2+r}K_{3/2-r}(L\sqrt{\mathcal{M}^2}|\vec{n}|),
\end{equation}
where $K_n(z)$ is the modified Bessel function of the second kind, and 
\begin{equation}\label{eq:sum}
\sum_{\vec{n}\neq 0} \equiv \sum_{n_x=-\infty}^\infty \sum_{n_y=-\infty}^\infty \sum_{n_z= -\infty}^\infty (1 - \delta(|\vec{n}|,0)),
\end{equation}
 with $\vec{n}=(n_x,n_y,n_z)$. It should be mentioned that the discrete summations in Eqs.~(\ref{eq:g1r},\ref{eq:deltar}) are only taken up to a certain number, $|n|_\mathrm{max}=\frac{L}{2a}$, where $L$ and $a$
are the lattice size and lattice spacing, respectively. Nowadays, most LQCD simulations adopt a $L/a$ in the range of $16\sim32$.

In the hybrid approach, the finite volume effect is calculated in the following way:
\begin{equation}\label{finite-volume-G}
  \delta G=\lim_{q_{max}\rightarrow\infty}\left[\frac{1}{L^3}\sum^{q_{max}}_{q_i}I(q_i)-\int^{q<q_{max}}\frac{d^3q}{(2\pi)^3}I(q)\right],
\end{equation}
where the function $I(q)$ is
\begin{equation}\label{kernel}
  I(q)=\frac{1}{2\omega(\vec{q})\omega'(\vec{q})}
  \frac{\omega(\vec{q})+\omega'(\vec{q})}{E^2-(\omega(\vec{q})+\omega'(\vec{q}))^2+i\epsilon},
\end{equation}
with $\vec{q}=\frac{2\pi}{L}\vec{n}$ ($\vec{n}\in \mathcal{Z}^3$), $\omega(\vec{q})=\sqrt{m^2+\vec{q}^2}$, $\omega'(\vec{q})=\sqrt{M^2+\vec{q}^2}$, and
$E=\sqrt{s}$.

In the L{\"u}scher method, the function $I(q)$ of Eq.~(\ref{kernel}) is reduced to~\cite{Doring:2011vk}
\begin{equation}\label{Luscher-I}
  I(q)=\frac{1}{2E}\frac{1}{p^2-q^2+i\epsilon},
\end{equation}
where $p=\lambda^{1/2}(E^2,M^2,m^2)/2E$. 

In the present paper we are also treating the $K^*$ as a stable particle, while in fact it has a width of around 45 MeV. In an unquenched calculation  if one uses interpolators
of $K^*\bar{K}$ one would reach the decay channels and one would have to deal with the three-body channels of $K\bar{K}\pi$. The formalism to deal with three body-systems in finite volume is also available in Refs.~\cite{  Polejaeva:2012ut,Briceno:2012rv,Hansen:2014eka}. For two-body systems with one unstable particle, one can use a formalism in which the self-energy of the unstable particle is discretized in the moving frame~\cite{Roca:2012rx}.  We shall not do this here, although when more refined lattice calculations are available it would be interesting to tackle this problem.
There are reasons not to do that at the present time. One of them is that many of the present lattice simulations use large pion masses where the decay channels would be blocked, but even there they
can assess the existence of a bound state of $K\bar{K}^*$ nature. The second reason is that in present lattice simulations, even using unquenched calculations, levels tied to channels that couple to certain quantum numbers do not show up unless explicit interpolators for this particular channel are explicitly used as interpolators. This was the case on the $\phi\rho$ system looking for the $a_1(1260)$ resonance
in  Refs.~\cite{Prelovsek:2011im,Lang:2014tia} and in the $KD$ system in Ref.~\cite{Torres:2014vna}, where the levels associated to the coupled channel $\eta D_s$ also did not show up in the
simulation. The reason for this fact seems to be that the coupled channels not considered would show up in the time evolution at times where noise appears in the simulations, preventing any signal from being seen. The argument has stronger weight for the decay channels of resonances with a small width, like the present one with $\Gamma=24$ MeV.
Obviously, there would be problems in the interpretation of the levels if these depend on the interpolators used, but the idea is to use interpolators with maximum overlap with the actual states, and there the effective field theories that we are using are of much help since they are telling the nature of the states under consideration. Then we suggest using interpolators that accommodate this structure, and in the present case these would be $K\bar{K}^*$ interpolators.

 It is true that the consideration of the decay channels of the particles involved in a problem leads to changes in the spectrum~\cite{Bernard:2007cm,Roca:2012rx,Briceno:2014uqa}
  and that to get the proper spectrum multihadron interpolators should be used~\cite{Dudek:2012xn}, but also, as mentioned in Ref.~\cite{Dudek:2012xn} , one can and must restrict oneself to lower energies if the interpolators accounting for the inelastic spectrum are not used. Concerning the present case we can use the analogy of this work, where we have $K\bar{K}^*$ and the $\bar{K}^*$ can decay to $\bar{K}\pi$ , and the case of Ref.~\cite{Roca:2012rx}, where one had $\rho\pi$ and the $\rho$ could decay to $\pi\pi$. In spite of the large width of the $\rho$, the first level was very similar in the analysis with  a stable $\rho$ or a decaying $\rho$.
 The second level changed a bit more in both approaches, but it is reasonable to expect that with a smaller width of the $K^*$, the differences would be much smaller. This, and other reasons that we will discus in Sec. IV concerning partial
 wave mixing, advise us to make use of only the first two levels that we shall discuss in the next section.

\section{Results and discussions}
\subsection{The energy levels}

\begin{figure}[htbp]
  \centering
  \includegraphics[angle=270,width=0.48\textwidth]{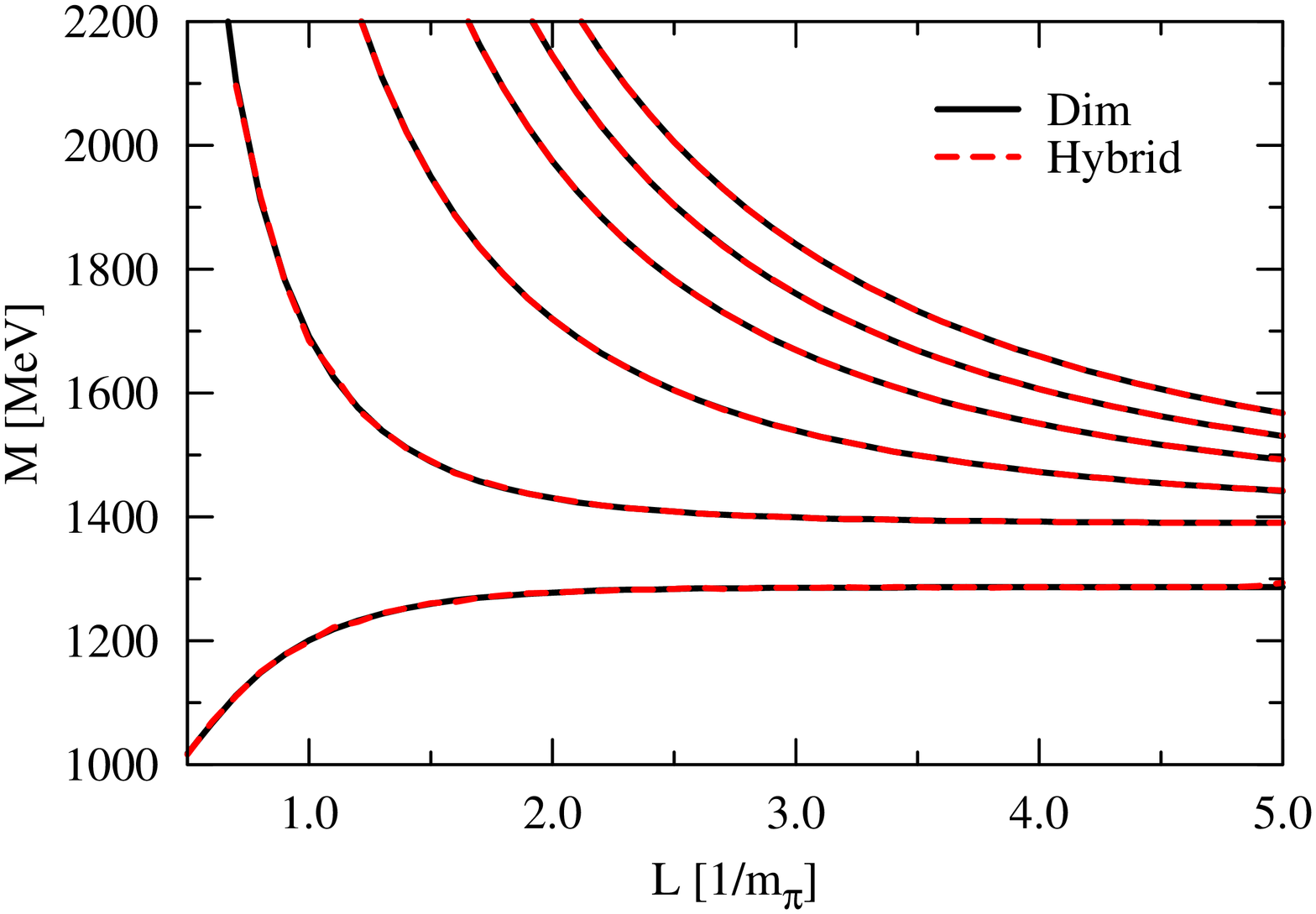}
   \includegraphics[angle=270,width=0.48\textwidth]{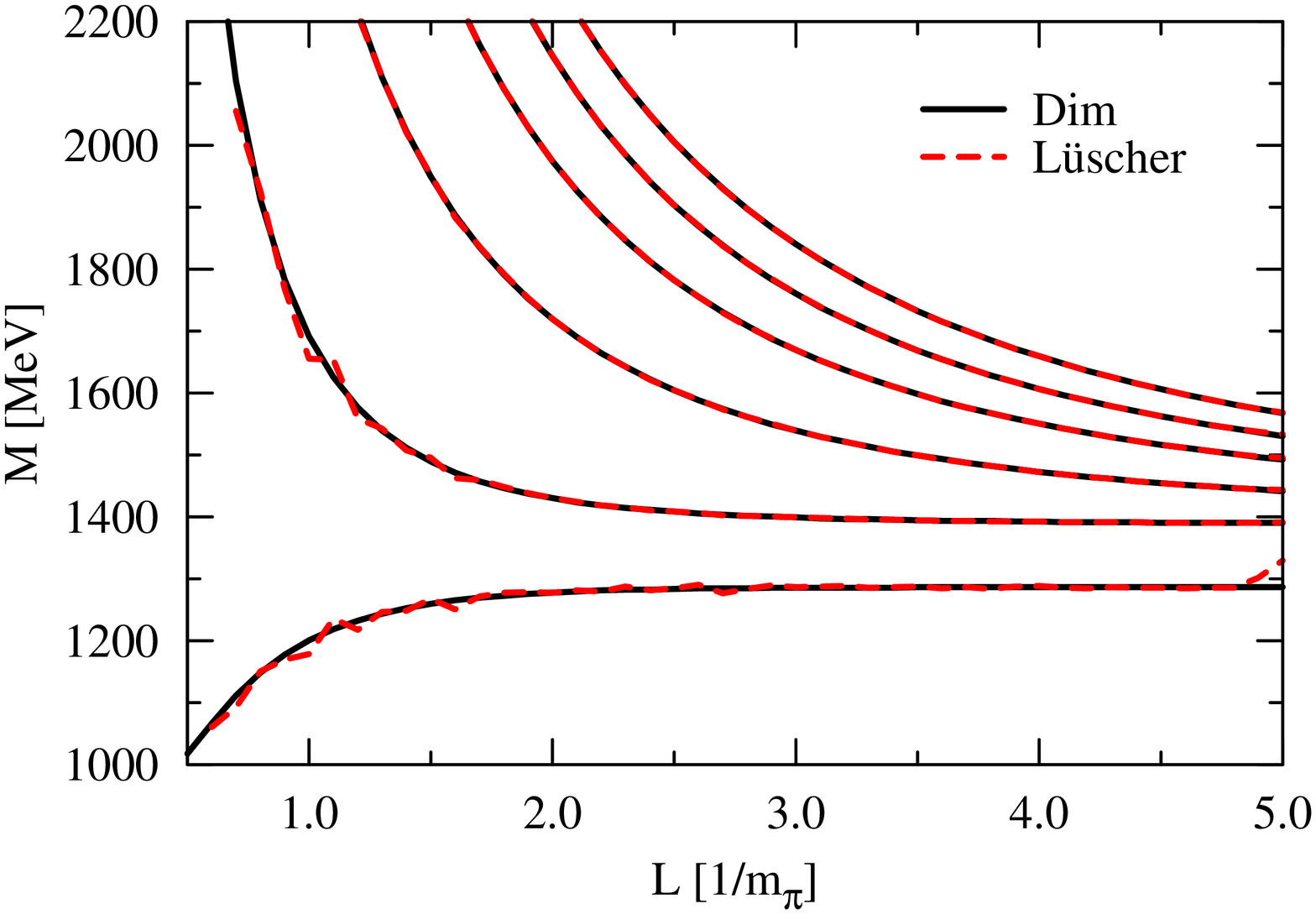}\\
  \caption{Energy levels of the $KK^*$ system with $\tilde{G}$ obtained from the dimensional regularization scheme in comparison with those obtained in the 
  hybrid approach (left) and the
 L{\"u}scher approach (right)   with $q_{max}=4000$ MeV. The lattice size $L$ is given in units of $1/m_\pi$, where $m_\pi$ is the
  physical pion mass.}
  \label{energylevels-a-qmax4000}
\end{figure}

\begin{figure}[htbp]
  \centering
  \includegraphics[angle=270,width=0.48\textwidth]{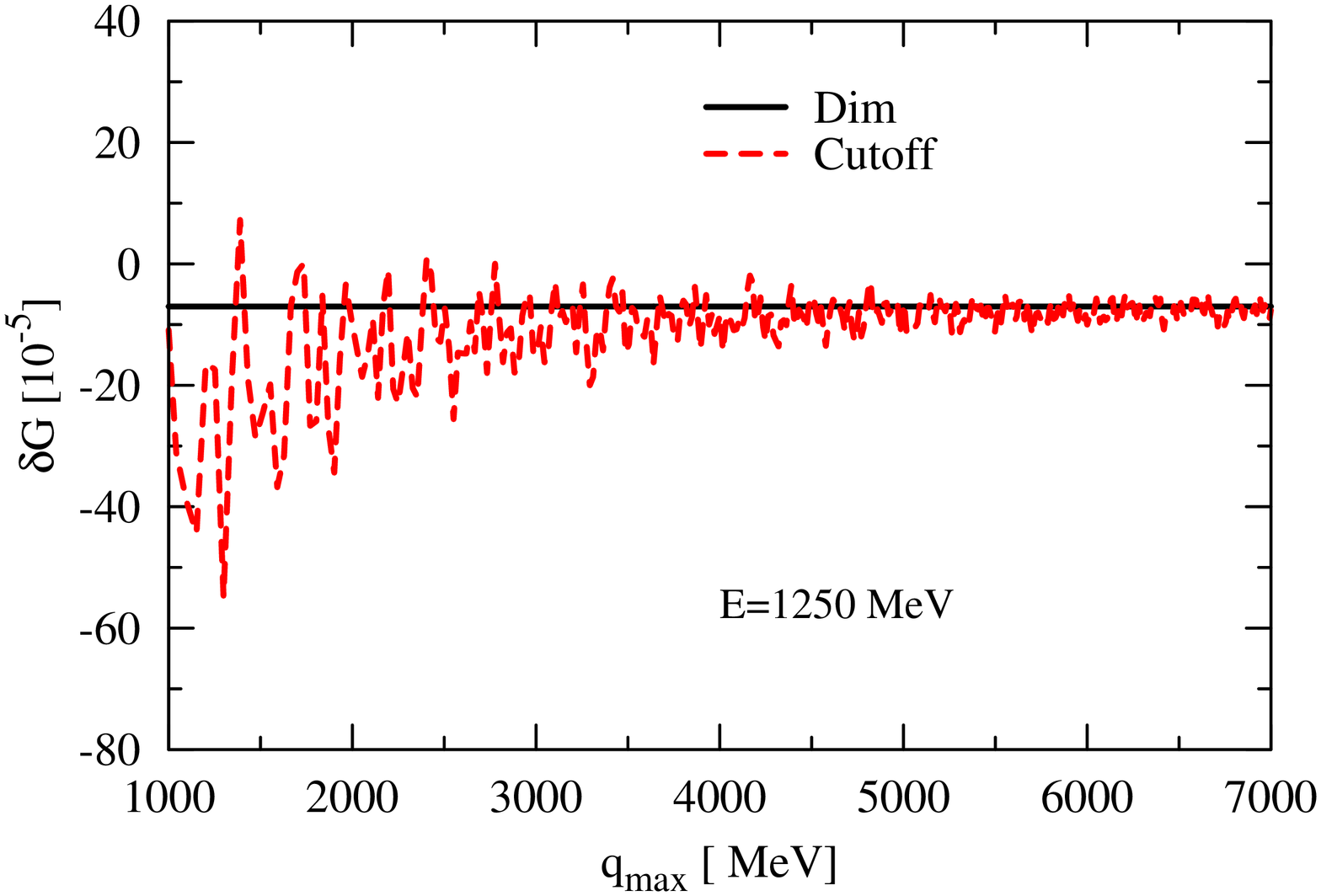}
    \includegraphics[angle=270,width=0.48\textwidth]{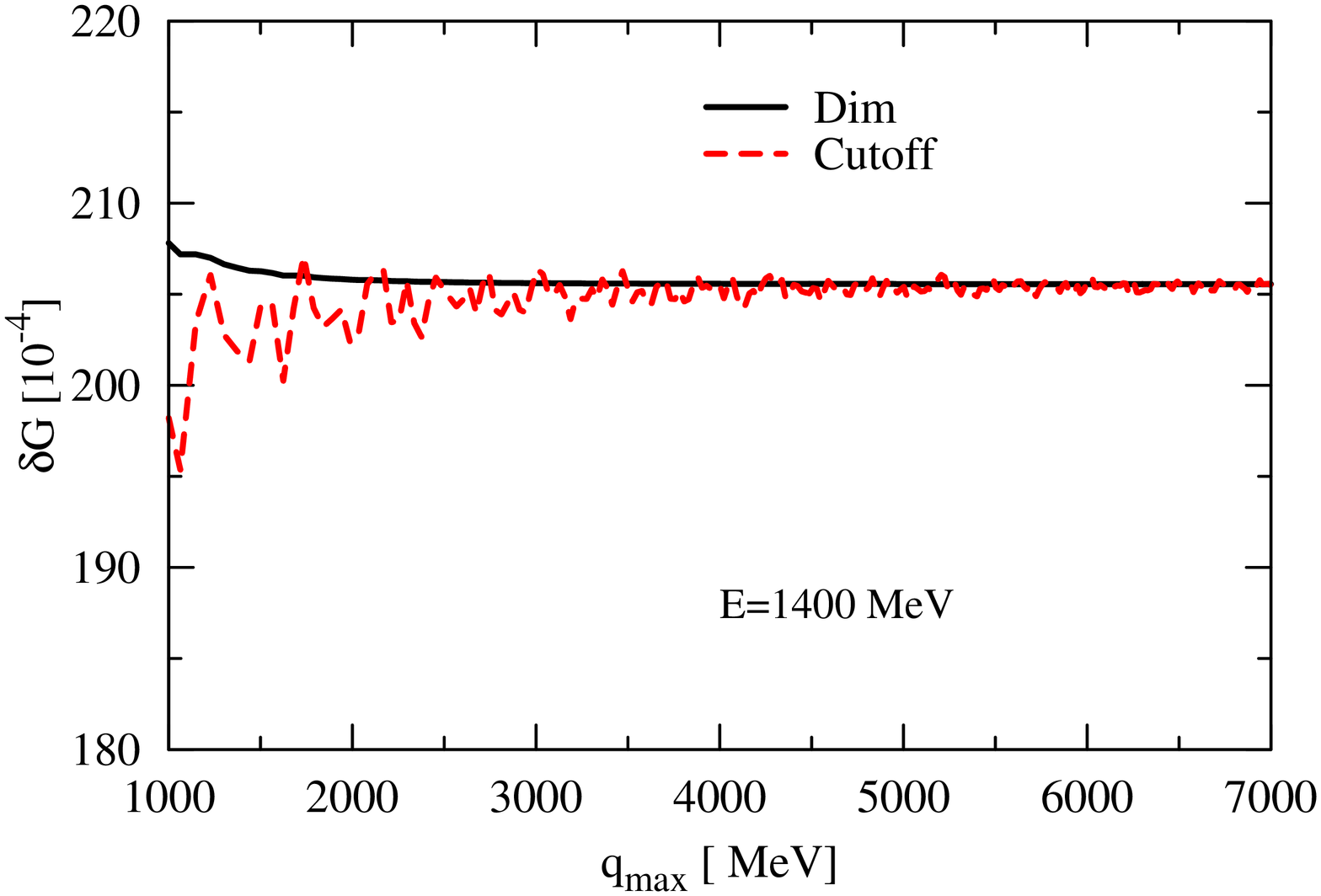}
  \caption{Finite-volume corrections, $\delta G$, for the $KK^*$ system calculated in the dimensional regularization scheme and the cutoff scheme with a sharp cutoff, $q_\mathrm{max}$, at
  the energy $E=1250$ MeV (left) and $E=1400$ MeV (right) and with $L=2.5/m_\pi$. }
  \label{fig:fluc}
\end{figure}

The left panel of Fig.~\ref{energylevels-a-qmax4000} shows the energy levels as functions of the cubic box size $L$ obtained in the dimensional regularization scheme. For the sake of comparison, we show as well
the energy levels obtained in the hybrid method with $q_{max}=4000$ MeV.  With the scale of Fig.~\ref{energylevels-a-qmax4000}, the two curves are hardly distinguishable. However, as noticed in 
all previous works, there are some unphysical discontinuities in the hybrid approach, which disappear with an average of the results obtained with several sharp cutoffs~\cite{MartinezTorres:2011pr} or with a smooth cutoff~\cite{Doring:2011vk}.   This can be better appreciated from Fig.~\ref{fig:fluc}, which shows that
the dimensional regularization scheme exhibits no sign of fluctuation, where small fluctuations can still be seen at a cutoff value of about 7000 MeV in the cutoff (hybrid) approach.\footnote{
In the dimensional regularization scheme, for the sake of comparison, $q_\mathrm{max}$ has been related to $|n|_\mathrm{max}$ via  $q_\mathrm{max}= \frac{2\pi}{L} |n|_\mathrm{max}$.} 
In the following, unless otherwise noticed, we work with the dimensional regularization method. 

The energy levels  obtained in the L{\"u}scher approach are shown on the right panel of Fig.~\ref{energylevels-a-qmax4000} as functions of the cubic box size $L$, in comparison with those obtained in the dimensional regularization scheme. It is clear that at least for the two lower energy levels,
the L\"{u}scher results show stronger fluctuations than those of the dimensional regularization approach  (also  than those of the hybrid approach).  Furthermore,
it is shown in Ref.~\cite{Zhou:2014ana} that the deduced phase shifts from the L\"{u}scher method can deviate by about 20 percent from the effective approach of Ref.~\cite{MartinezTorres:2011pr} at the energy region where the resonance dominates, at least for the $\pi K$ interaction in the $K^*$ channel (see Fig. 12 of Ref.~\cite{Zhou:2014ana}).

As discussed in Ref.~\cite{Doring:2011vk}, the new terms incorporated in Ref.~\cite{Doring:2011vk} with respect to the L\"{u}scher approach are exponentially suppressed and one would wonder whether other exponentially suppressed contributions from $t$ and $u$ channels, neglected in both approaches are not equally relevant. In this sense, explicit calculations of these effects done for mesons in the scalar sector~\cite{  Albaladejo:2012jr}, or the vector sector~\cite{ Albaladejo:2013bra}, show them to be negligible for lattice sizes bigger than $L=1.5$ $m_\pi^{-1}$.

The $a_1(1260)$ and $b_1(1235)$ states have recently been studied in $N_f=2$ lattice QCD~\cite{Lang:2014tia}, where in addition to $q\bar{q}$ interpolators, 
meson-meson interpolators were also taken into account.  
Compared with the $a_1(1260)$ and $b_1(1235)$, the $f_1(1285)$ is more suited to test the dynamical nature of the axial-vector mesons because of the following reasons.
First, it is a single channel problem. Second, it is a bound state. Therefore it appears as a discrete energy level even in LQCD simulations. Third, it is built from
the interaction of two strange mesons, which makes it less susceptible to chiral extrapolations. 

The ground-state pseudoscalar mesons and vector mesons have been studied in a number of $n_f=2+1$ LQCD simulations~\cite{Durr:2008zz,Aoki:2008sm,WalkerLoud:2008bp,Lin:2008pr,Bietenholz:2011qq}.  Some of the gauge configurations are available on the 
International Lattice Data Grid, e.g., the PACS-CS configurations~\cite{Aoki:2008sm}, which in principle makes a study of the $f_1(1285)$ straightforward.  In Table \ref{tab:pacs-cs}, we show
the masses of the $f_1(1285)$ calculated in our framework, defined as  the energies where $\tilde{T}$ has a pole below the $KK^*$ threshold, with the $K$, $K^*$ masses,  and the lattice size $L$ of the PACS-CS configuration~\cite{Aoki:2008sm} (note, however, that these masses are calculated there with only $q \bar q$ interpolators).  It is interesting to note that
the $f_1(1285)$ remains as a bound state at these unphysical situations and the binding energy increases as the masses of its components
increase. 

\begin{table}[htbp]
  \caption{Masses, $M$,  and binding energies, $B$,  of the $f_1(1285)$ at unphysical quark masses and in finite volume. The $K$ and $K^*$ masses are
  those obtained by the PACS-CS $n_f=2+1$ simulations~\cite{Aoki:2008sm}.  In the last row, the numbers in the parentheses  are the uncertainties
  coming from those of the $K^*$ and $K$ added in quadrature.  All the energies are in units of MeV while the lattice size $L$ is in units of fm.}\label{tab:pacs-cs}
  \centering
  \begin{tabular}{l|cccccc|c}
    \hline\hline
     Inputs & Conf1& Conf2 & Conf3 & Conf4 & Conf5 & Conf6 & Physical \\
    \hline
 $m_K$  &   554(8) & 594(9) & 582(9) & 635(9) & 713(10) & 789(11) &  495.0\\
 $M_{K^*}$  & 939(17) & 984(16) & 963(16) & 1015(15) & 1078(17) & 1156(17) & 893.1\\
 $L$ & 2.90(4) & 2.90(4) & 2.90(4) & 2.90(4) & 2.90(4) & 2.90(4) & $\infty$\\
  $M$ & 1367  & 1442 &  1412 &  1506&1635 &1785&  1286\\
  $B$  & 126(19) & 136(18) & 133(18) & 144(17) & 156(20) & 160(20) & 102.1\\
    \hline\hline
  \end{tabular}
\end{table}

\begin{figure}[htbp]
  \centering
  \includegraphics[angle=270,width=0.65\textwidth]{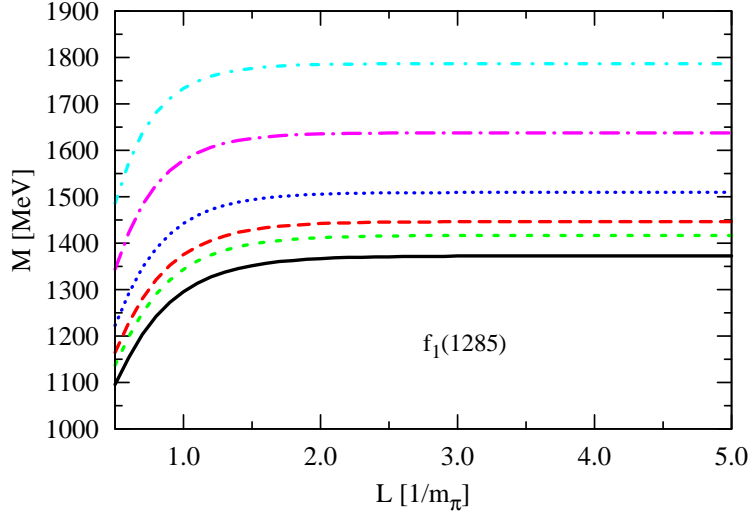}\\
  \caption{Mass of the $f_1(1285)$ as a function of the lattice size $L$ at  six combinations of light and strange quark masses, from lower to upper corresponding to
  Conf1 to Conf6 of Table \ref{tab:pacs-cs}, respectively.}
  \label{fig:f1}
\end{figure}

In Fig.~\ref{fig:f1}, we show the mass of the $f_1(1285)$ as a function of the
lattice size $L$ at six different combinations of light and strange quark masses, corresponding to those of the PACS-CS configurations. 
It is clear that  the results already approach their continuum limits at a lattice size of two to three times $1/m_\pi$.

\subsection{The inverse problem}
In this section we tackle the inverse problem of extracting an effective potential from discrete energy levels of LQCD. Close to the
$KK^*$ threshold, one can assume a potential of the following form:
\begin{equation}\label{potential}
  V=a+b[s-(M_{K^*}+m_K)^2].
\end{equation}
The two parameters $a$ and $b$ can be determined by fitting to the lattice energies.

We assume that the first and second energy levels shown in Fig.~\ref{energylevels-a-qmax4000} are "LQCD" data. We take three energies from the first level and three more from the second one, and assign them an error of $10$ MeV. The corresponding values of $L$ are: $2.0\,m_\pi^{-1}, 3.0\,m_\pi^{-1}, 4.0\,m_\pi^{-1}$. 
Performing a least-squares fit, we obtain a $\chi_\mathrm{min}^2\approx1\times10^{-5}$ and  the following  two values:
\begin{equation}\label{parameters}
  a=-157\pm29,\quad b=(-1.4\pm1.1)\times10^{-4} \mbox{ MeV}^{-2}
\end{equation}
With the potential of Eq.~(\ref{potential}), by solving the corresponding Bethe-Salpeter equation, one finds a bound state at $M=1286\pm 37$ MeV, whose central value coincides with
the original  value we started with.  It should be noted that although the bound state approaches its continuum limit as $L$ increases,
the potential approach has the advantage that it can connect the LQCD energy levels
at moderate $L$ or small $L$ with the binding energies in the continuum in a quite accurate and model-independent way (for a relevant and extensive discussion, see, e.g., Ref.~\cite{Albaladejo:2013aka}).

 Of course, for the case at hand, one does not need to go through the inverse process to obtain the $f_1(1285)$ because it appears as a  bound state.  Nevertheless,
this procedure allows us to obtain an effective potential in a more or less model-independent way. 

Following the approach of Refs.~\cite{Hyodo:2013nka,Aceti:2014ala,Sekihara:2014kya},  one can quantify  the relative contributions of the meson-meson component in the 
$f_1(1205)$ wave function. The coupling constant of a resonance to its component channel can be calculated as follows,
\begin{equation}
g^2=\lim_{s\rightarrow s_0} (s-s_0)T=\lim_{s\rightarrow s_0} \frac{s-s_0}{V^{-1}-G}=\left.\frac{1}{\frac{\partial V^{-1}}{s}-\frac{\partial G}{\partial s}}\right|_{s=s_0},
\end{equation}
where $s_0$ is the pole position. From the above equation, one can obtain  the  identity,
\begin{equation}
-g^2\frac{\partial G}{\partial s} + g^2\frac{\partial V^{-1}}{\partial s}=1.
\end{equation}
The first term gives the contribution of the composite component being dynamically generated, while the second term gives the rest (e.g., genuine $q\bar{q}$ or missing meson-meson channels). For the $f_1(1285)$, we find that $-g^2\frac{\partial G}{\partial s}=0.50$, which implies that the meson-meson component accounts for about half of the $f_1(1285)$ wave function.  Given the fact that the $f_1(1285)$ is located about 100 MeV below the $KK^*$ threshold, this value does not seem that small.

\section{$D$-waves for $K\bar{K}^*$ with chiral Lagrangians}
In dealing with the $K\bar{K}^*$ system with finite volumes one has to look at spin projections and partial wave mixing. There has been much work done along these lines recently.  By using $q\bar{q}$ interpolators~\cite{ Thomas:2011rh,Edwards:2011jj,Dudek:2009qf,Dudek:2010wm} new methods and suitable interpolators have been developed to project on the desired spin states. More relevant to our problem, using the L\"{u}scher formalism for scattering of two particles, several papers have dealt with this problem. A detailed study for the case of $0^-$, $1/2^+$ interacting particles is done in Ref.\cite{  Bernard:2008ax}, which is generalized in Ref.~\cite{Gockeler:2012yj} to moving frames. The case of $1/2^+$, $1/2^+$ interacting particles is studied in Ref.\cite{Briceno:2013lba}~\footnote{See Refs.~\cite{  Li:2012bi,Li:2014wga} for the generalized L\"{u}scher formula in multichannel meson(baryon)-baryon scattering formulated in both non-relativistic quantum mechanics and quantum field theory.}
 and applied to the deuteron case in Ref.~\cite{Briceno:2013bda}. A formal extension to the scattering of particles with arbitrary spin is done in Ref.~\cite{Briceno:2014oea}. 
 For a first study of coupled-channel effects in LQCD simulations,  see, e.g., Refs.~\cite{  Dudek:2014qha,Wilson:2014cna}. 
 
In the present case, we are concerned about the scattering of $0^-$ and $1^-$ particles in the rest frame of the particles.  If we only took into account $S$-wave interaction between the $0^-$ and
			$1^-$ particles, it can be shown that on the cubic lattice and for total momentum $\vec{P}=0$ the $S$-wave only mixes with the $G$-wave. However,  in infinite volume, the $f_1(1285)$ can decay into a pair of
 $1^-$ and $0^-$ particles via the $D$-wave. In this case, mixing of $L=0$ and $L=2$ can occur. In order to assess the relevance of this component in the problem that we study we go back to the theory that generates the interaction of these particles using chiral dynamics. The chiral Lagrangian for this interaction is given in 
Ref.~\cite{Birse:1996hd} by 
\begin{equation}
\mathcal{L}_\mathrm{VVPP}=-\frac{1}{4f^2}\mathrm{Tr}([V^\mu,\partial_\nu V_\mu][P,\partial^\nu P]),
\end{equation}
which leads to the potential 
\begin{equation}
\tilde{V}\sim (p_1+p_3)(p_2+p_4)\epsilon^\mu\epsilon_\mu,
\end{equation}
where $p_1$, $p_2$, $p_3$, $p_4$ correspond to the two incoming and two outgoing momenta  in $K\bar{K}^*\rightarrow K\bar{K}^*$, and $\epsilon^\mu$ is the polarization of the vector. It is clear that this potential has no $D$-waves. However, $D$-waves are automatically generated where this Lagrangian is reinterpreted by means of the local hidden gauge approach ~\cite{ Bando:1987br,Harada:2003jx,Meissner:1987ge} and
is generated by the exchange of a light vector meson ($\rho $ meson for  the $K\bar{K}^*$ interaction). In this case one has the explicit $\rho$ propagator and the potential becomes 
\begin{equation}
\tilde{V}\sim\frac{(p_1+p_3)\cdot(p_2+p_4)\epsilon^\mu \epsilon_\mu}{-(\vec{p_1}-\vec{p_3})^2-m_\rho^2},
\end{equation}
which this time develops a $D$-wave. It is easy to see that the ratio of the $D$-wave to $S$-wave is (using $|\vec{p_1}-\vec{p_3}|<m_\rho$ for the derivation)
\begin{equation}
\frac{\tilde{V}_2}{\tilde{V}_0}\approx\frac{2}{3}\frac{\vec{p}\,^4}{m_\rho^2}\frac{1}{E_1E_2+\vec{p}\,^2/2}
\end{equation}
with $\vec{p}$ the CM momentum.

Now we look at the energy levels of Fig.~1, that we have used for the simulation. We actually took the first two levels for $L m_\pi>2$ in the inverse analysis. Then
we consider the levels $2$, $3$, $4$ that have energies in the continuum and we find the ratios for $Lm_\pi=2$
\begin{eqnarray}
\frac{\tilde{V}_2}{\tilde{V}_0}&=&0.002\quad\quad\mbox{for level 2},\\ 
                           &=&0.079\quad\quad\mbox{for level 3},\\
                            &=&0.208 \quad\quad\mbox{for level 4}.
\end{eqnarray}
The numbers would be further reduced by Clebsh-Gordan coefficients of $L=2$ and $S=1$ to give $J=1$, which are unity for $L=0$. For bigger values of $L$, these energies are smaller and these ratios also decrease. For instance for $Lm_\pi=3$ and level 4 we would find $\tilde{V}_2/\tilde{V}_0\approx0.059$.

The discussions conducted here are useful, because since we have only used the levels 1 and 2 for $Lm_\pi\ge2$, then we always have a ratio of $\tilde{V}_2/\tilde{V}_0$ smaller than two per thousand, and we can safely ignore the mixing. However, we also see that if we were to use the level 4 in our analysis and for values of $Lm_\pi< 2$ we would have mixture of the order of 25\% which would require us to explicitly consider the mixing for a proper interpretation of the results.

The $D$-wave decay of the $f_1(1285)$ could in principle induce more
complicated mixing patterns. For spin-0 and spin-1 scattering, which is the
present case, it can be shown that the $J=1$ $D$-wave of the $f_1(1285)$
does not mix with any of the various $P$- and $F$-waves. However, it mixes
with the $J=3$ $D$-wave. Indeed, at energies much higher than considered
here, $J=3$ resonances have been found that decay into $KK^*$, such as
the $\phi_3(1850)$~\cite{Agashe:2014kda}. 

However, for the purpose of a rough error estimate, we can assume that
the $J=3$ $D$-wave is of a similar size as the $D$-wave induced by Eq.~(24),
such that the uncertainties quoted in Eqs.~(26)-(28) might be larger by a
factor of 2. If lattice data become more precise, a coupled-channel
calculation including the $S$-wave and the two $D$-waves will be
necessary. At nonzero total momentum, which is not considered here, the
mixing can become more complicated~\cite{Gockeler:2012yj}.

\section{Summary}
We have studied the $KK^*$ interaction in the $f_1(1285)$ channel in finite volume with the  chiral unitary approach. The relativistic loop function was calculated in
the dimensional regularization scheme and compared with the hybrid approach developed previously. It was shown that although both approaches yield the same results if treated properly, the dimensional regularization scheme is numerically more stable. 
In addition, we found that the L{\"u}scher method fluctuates more strongly with the variation of the cutoff, but agrees with the hybrid method qualitatively. 

In anticipation of future lattice QCD studies, we have calculated the position of the $f_1(1285)$ at six different combinations of light and strange quark masses as a function of the lattice size $L$.  If confirmed, this could provide another test of the $f_1(1285)$ being a dynamically generated state. Indeed, the $KK^*$ meson-meson component is found to  account for one half of its wave function.

\section{Acknowledgements}

 We thank A. Rusetsky and R. Briceno for providing us useful information  and  Michael D\"oring for valuable discussions on the partial wave mixing in finite volume. 
 One of us, E. O., wishes to acknowledge support from
the Chinese Academy of Sciences (CAS) in the Program of
Visiting Professorship for Senior International Scientist. X.L.R. acknowledges the financial support from the China Scholarship Council and support from the Innovation Foundation of Beihang University for Ph.D. Graduates.
 This work is supported in part by the National Natural Science Foundation of China under Grants No. 11375024, No. 11205011, and No. 11105126; the New Century Excellent Talents in University Program of Ministry of Education of China under Grant No. NCET-10-0029; the Fundamental Research Funds for the Central Universities; the Spanish Ministerio
de Economia y Competitividad and European FEDER
funds under Contracts  No. FIS2011-28853-C02-01
and No. FIS2011-28853-C02-02; the Generalitat Valenciana
in the program Prometeo II-2014/068; and the European Community Research
Infrastructure Integrating Activity Study of
Strongly Interacting Matter (HadronPhysics3,
Grant No. 283286) under the Seventh Framework
Programme of EU.

\end{document}